\documentclass[10pt,letterpaper]{article}

\usepackage{textcomp}
 \usepackage{times}
\usepackage{arxiv}

\usepackage[utf8]{inputenc} 
\usepackage[T1]{fontenc}    
\usepackage{hyperref}       
\usepackage{url}            
\usepackage{booktabs}       
\usepackage{amsfonts}       
\usepackage{nicefrac}       
\usepackage{microtype}      
\usepackage{lipsum}
\usepackage{textgreek}
\usepackage[squaren]{SIunits}

\usepackage[utf8]{inputenc}

\usepackage{cite}

\usepackage{nameref,hyperref}

\usepackage[right]{lineno}

\usepackage{microtype}
\DisableLigatures[f]{encoding = *, family = * }

\raggedright
\usepackage{changepage}

\usepackage[aboveskip=1pt,labelfont=bf,labelsep=period,singlelinecheck=off]{caption}

\makeatletter
\renewcommand{\@biblabel}[1]{\quad#1.}
\makeatother

\usepackage{lastpage,fancyhdr,graphicx}
\usepackage{epstopdf}
\fancyheadoffset[L]{2.25in}
\fancyfootoffset[L]{2.25in}

\usepackage{color}

\definecolor{Gray}{gray}{.25}

\usepackage{graphicx}

\usepackage{sidecap}

\usepackage{wrapfig}
\usepackage[pscoord]{eso-pic}
\usepackage[fulladjust]{marginnote}
\reversemarginpar

\title{Megahertz X-ray microscopy at X-ray Free-Electron Laser and Synchrotron sources}

\begin{document}
\maketitle
\begin{flushleft}

Patrik Vagovi\v{c}\textsuperscript{1,2,3,*},
Tokushi Sato\textsuperscript{1,2},
Ladislav Mike\v{s}\textsuperscript{2},
Grant Mills\textsuperscript{2},
Rita Graceffa\textsuperscript{2},
Frans Mattsson\textsuperscript{4},
Pablo Villanueva-Perez\textsuperscript{4,1},
Alexey Ershov\textsuperscript{5},
Tom\'{a}\v{s} Farag\'{o}\textsuperscript{5},
Jozef Uli\v{c}n\'y\textsuperscript{6},
Henry Kirkwood\textsuperscript{2},
Romain Letrun\textsuperscript{2},
Rajmund Mokso\textsuperscript{4},
Marie-Christine Zdora\textsuperscript{7,8},
Margie P. Olbinado\textsuperscript{9},
Alexander Rack\textsuperscript{9},
Tilo Baumbach\textsuperscript{5},
Alke Meents\textsuperscript{1},
Henry N. Chapman\textsuperscript{1},
Adrian P. Mancuso\textsuperscript{2,10},
\\
\end{flushleft}
\bigskip
\begin{centering}
$^1$\it{Center for Free-Electron Laser, Notkestraße 85, 22607 Hamburg, Germany}
\\
$^2${European XFEL, Holzkoppel 4, 22869 Schenefeld, Germany}
\\
$^3${Institute of Physics, Academy of Sciences of the Czech Republic v.v.i., Na Slovance 2, 182 21, Praha 8, Czech Republic}
\\
$^6${ Faculty of Science, Department of Biophysics, P. J. Šafárik University, Jesenná 5, 04154 Košice, Slovakia}
\\
$^4${Lund University, Sweden}
\\
$^7${Diamond Light Source, Harwell Science and Innovation Campus, Didcot, Oxfordshire OX11 0DE, United Kingdom}
\\
$^8${Department of Physics \& Astronomy, University College London, London, United Kingdom}
\\
$^5${Institute for Photon Science and Synchrotron Radiation, Karlsruhe Institute of Technology (KIT), Herrmann-von-Helmholtz-Platz 1, 76344 Eggenstein-Leopoldshafen, Germany.}
\\
$^9${ESRF -- The European Synchrotron, 71 Avenue des Martyrs, 38000 Grenoble, France}
\\
$^1${Department of Chemistry and Physics, La Trobe Institute for Molecular Science, La Trobe University, Melbourne, Victoria, 3086, Australia}
\\
\end{centering}
\bigskip
* patrik.vagovic@cfel.de

\begin{abstract}
We demonstrate X-ray phase contrast microscopy performed at the European X-ray Free-Electron Laser sampled at 1.128 MHz rate. We have applied this method to image stochastic processes induced by an optical laser incident on water-filled capillaries with micrometer scale spatial resolution. The generated high speed water jet, cavitation formation and annihilation in water and glass, as well as glass explosions are observed. The comparison between XFEL and previous synchrotron MHz microscopy shows the superior contrast and spatial resolution at the XFEL over the synchrotron. This work opens up new possibilities for the characterization of dynamic stochastic systems on nanosecond to microsecond time scales at megahertz rate with object velocities up to few kilometers per second using X-ray Free-Electron Laser sources. 
\end{abstract}

\begin{centering}
\begin{figure*}[!htbp]
 \includegraphics[width = 1\textwidth]{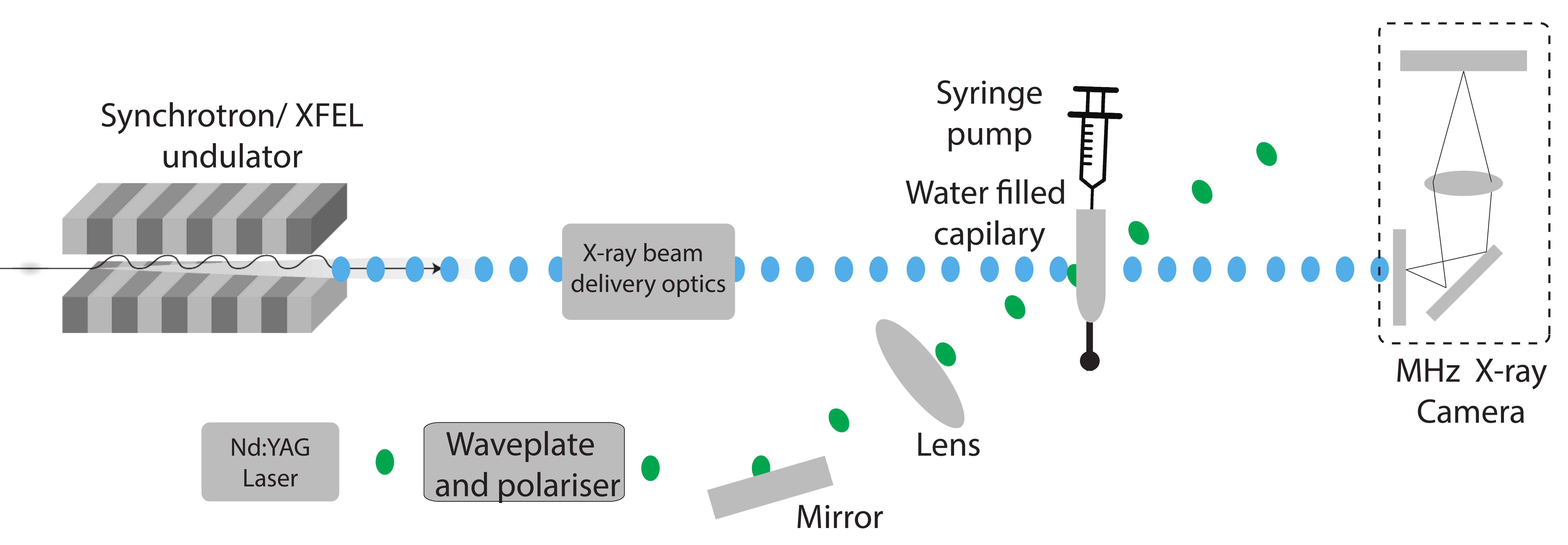}
\caption{Schematics of the time-resolved MHz X-ray microscopy of laser induced dynamics in a water-filled glass capillary.} 
\label{fig:setup}
\end{figure*}
\end{centering}

Hard X-ray beams are well suited for microscopic 2D and 3D imaging of samples not transparent to visible light due to their high penetration power. Over the last two decades the field of X-ray imaging has developed considerably, mainly due to the availability of modern, third-generation synchrotrons producing X-rays of high brilliance \cite{Kunz:2001}. These sources have provided access to the structural determination of specimens down to nano meter scale resolutions. Exploiting the (partial) spatial coherence of synchrotron X-ray probes, several phase sensitive techniques have been developed providing access to the electron density of specimens either via X-ray optical analyzers \cite{Bonse:65, Chapman:1997,David:2002} or sophisticated algorithms \cite{Paganin:2002, Rodenburg:2004}.
While much attention has been paid to improve the spatial resolution of X-ray imaging to its limits, fewer resources have been used to explore the boundaries of the temporal domain. With the progress in the development of detectors over the last decade \cite{Hatsui:2015}, fast radiography and tomography with kilohertz frame rates are available allowing, for example, \texttildelow 100 tomograms per second \cite{Mokso:2017,Yashiro:2017}. 
Only relatively recently has the stroboscopic nature of synchrotrons been exploited. For example, imaging with synchronized or individual X-ray pulses applied to fast stochastic transient processes \cite{Fezzaa:2008, Olbinado:2017,Parab:2018}. The new paradigm of ultra-fast X-ray imaging could be introduced by megahertz X-ray Free-Electron Laser sources, where the high flux per pulse can reveal dynamics of stochastic processes with velocities up to the scale of several km/s with sub-micron scale resolutions with high sensitivity to projected densities.    
In this work we exploit the unique properties of the first operational hard X-ray megahertz XFEL source European XFEL (EuXFEL) and explore its possibilities for ultra-fast X-ray microscopy with megahertz sampling. The laser induced dynamic processes in an open ended glass capillary filled with water was used as a dynamic sample. We use this simple model system to show the advantages of microsecond temporal resolution, micrometer spatial resolution, ultra-short exposure below 100 fs and the improved signal-to-noise in the images all brought about by using a megahertz repetition rate XFEL. We compare the results obtained at EuXFEL to that at ESRF ID19 beamline with the setup depicted in Fig. \ref{fig:setup}. The first results obtained at EuXFEL are already comparable to the state of the art synchrotron performance \cite{Parab:2018} and are still subject to improvement.
Further developments will allow for improving the spatial resolution beyond the reach of most brilliant synchrotron sources with potential acquisition of 3D megahertz movies by employing X-ray multi-projections imaging~\cite{Villanueva:2018}.   

It has been reported that liquid jets with velocities as high as 850 m/s can be generated by focusing a visible light laser into a capillary filled with water \cite{Tagawa:2012}. A jet based on this principle may be considered for needle-free drug-delivery injection or for sample delivery applications at megahertz XFEL sources as a jet on demand which can significantly minimize the sample usage. The characterization of the laser-induced jets and fluidics  on the \unit{\nano\second} to \unit{\micro\second} time  scales is typically done by visible light microscopy. However, details on the microstructure are difficult or impossible to access with light microscopy due to the large refraction angle and strong multiple light scattering at the interfaces caused, for example, by micro cavitations. 
To explore the dynamics induced by the focused frequency-doubled Nd:YAG laser (Minilite II, Continuum) in the glass capillaries filled with water, we constructed conceptually similar time-resolved microscopy setups at the ID19 beamline at ESRF and at the SPB/SFX instrument at EuXFEL using an indirect scintillator-based detector coupled to the MHz FT-CMOS camera SHIMADZU HPV-X2 schematically depicted in Fig. \ref{fig:setup}. 

For the experiment at EuXFEL we used the SPB/SFX instrument \cite{Mancuso:2019}. For the X-ray microscopy measurements, we re-used the spent beam from the upstream interaction region and was out-coupled into air via a 180 \unit{\micro\metre} thick diamond window. 
A photon energy of 9.3 keV was used and the pulse train was filled with 128 X-ray pulses with a repetition rate of 1.128 MHz. The effective pixel size of the imaging system was \unit{3.2}{\micro\metre} (see Supplement 1). 

\begin{figure*}[!htbp]
 \includegraphics[width = 1\textwidth]{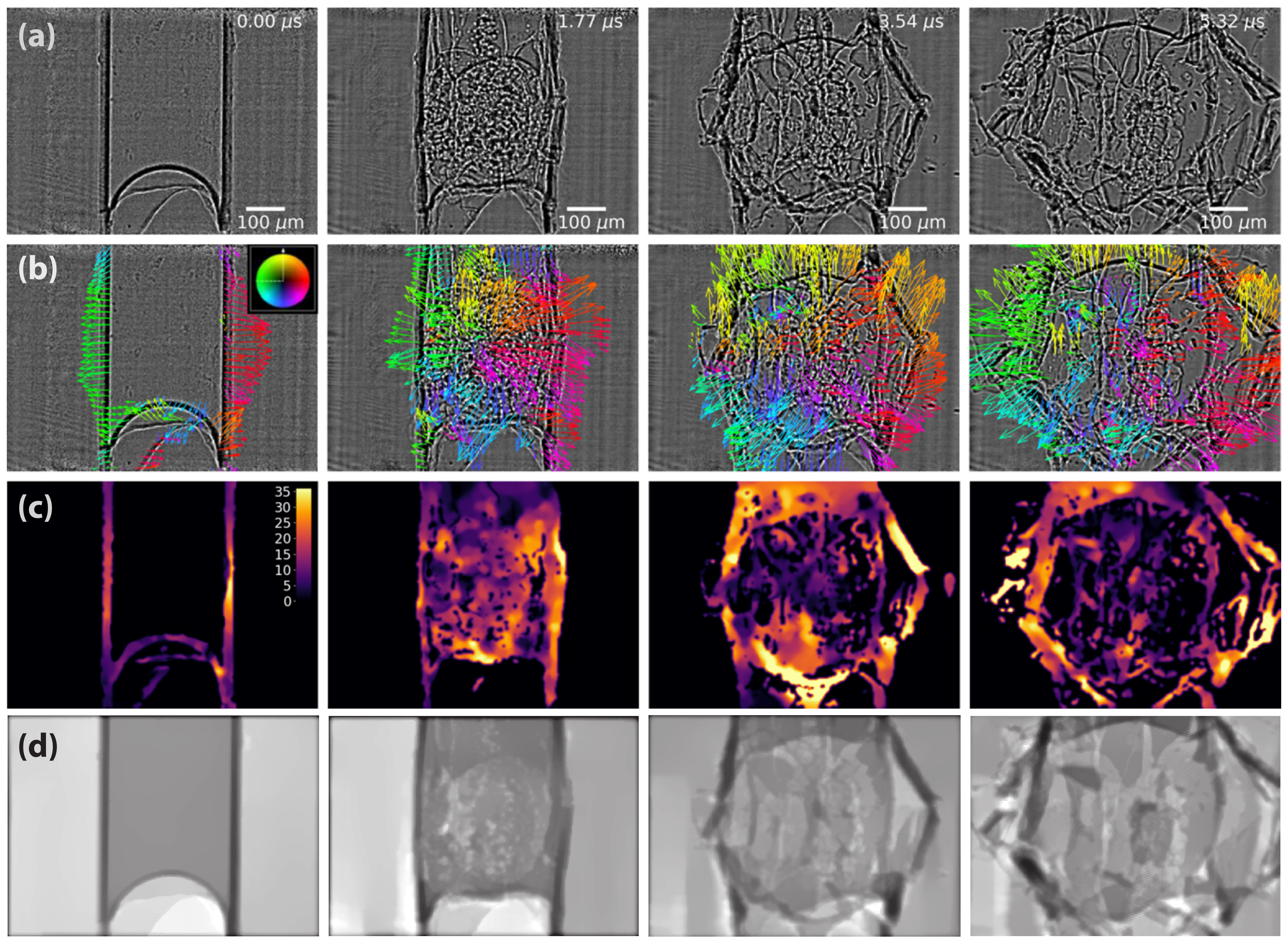}%
\caption{Image sequence of laser-driven explosion of {a} capillary filled with water imaged at EuXFEL. Sequence (a) is the result of high-pass adaptive filtering to remove the high-frequency noise and image flickering, sequences (b) and (c) are the result of optical flow analysis \cite{Myagotin12} shown as a directional vector for the movement of debris (b) and the velocity maps (c). The phase retrieval of the corresponding sequence (d) is performed using an ADMM-CTF algorithm \cite{Pablo:2017}.}
\label{fig:xfel_results}
\end{figure*}
For the experiment at the ESRF synchrotron we used the 16-bunch filling mode providing a bunch separation of 176 ns and the MHz camera was recording every third frame with an inter-frame time of 530 ns. The harmonics of the undulator with central photon energy 32 keV were conditioned by a set of 1D compound refractive lenses to enhance the flux density at the detector. The effective pixel size of the imaging system was \unit{8}{\micro\metre} (see Supplement 1).
\begin{figure*}[!htbp]
 \includegraphics[width = 1\textwidth]{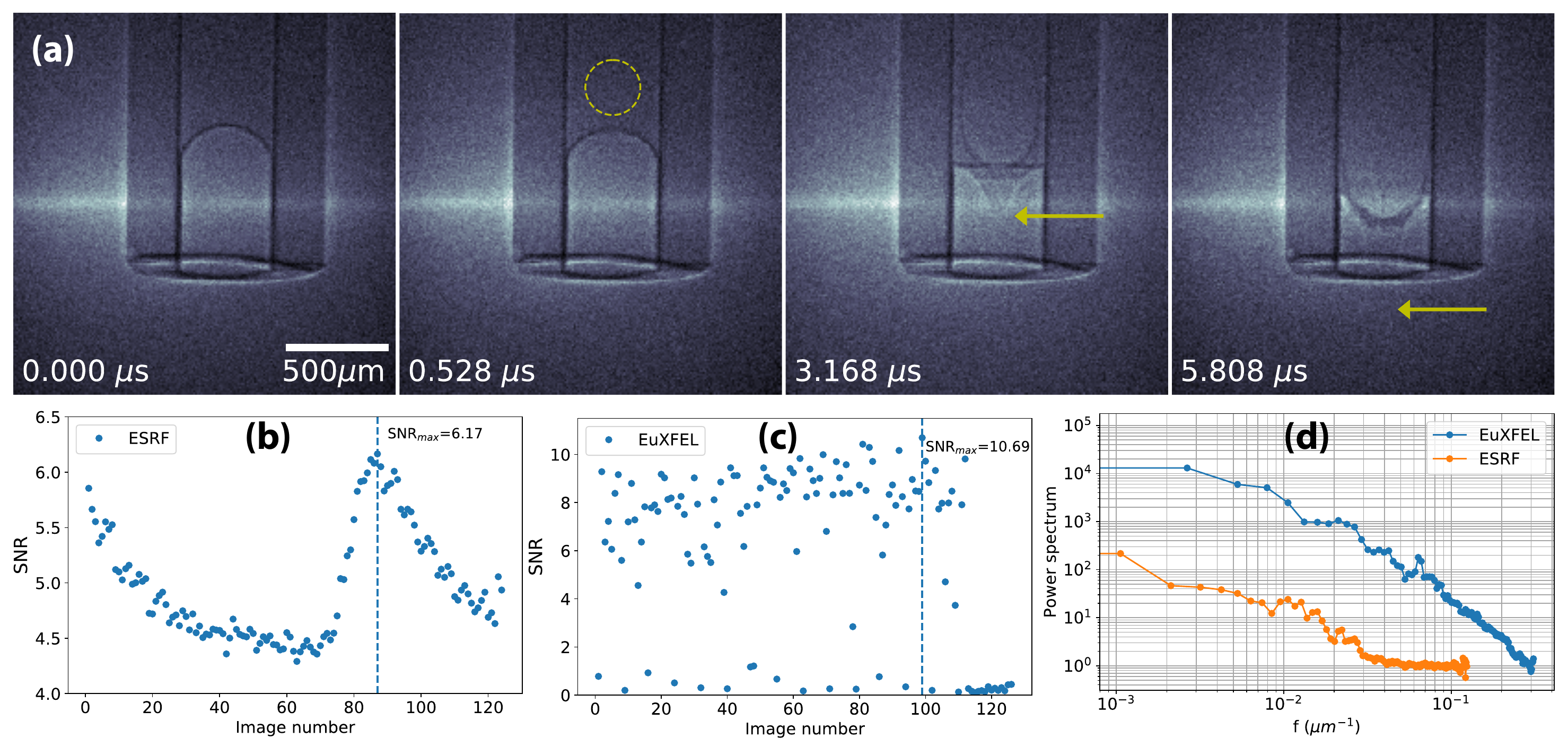}%
\caption{Image sequence of a water jet generated by absorbed power (\texttildelow 2 mJ pulse energy) of a focused visible-light laser inside the glass capillary imaged at the ESRF synchrotron (a). The circle on the second frame shows the laser-induced cavitation with radius of 144 \textmu m indicating an initial velocity of 272 m/s of the expanding wall. The arrows on the third and fourth frame indicate the tip of the water jet reaching a velocity of 184 m/s. The signal to noise analysis for ESRF (b) and EuXFEL (c) and power spectrum curves (d) shows superior image performance of EuXFEL data.}

\label{fig:esrf_results}
\end{figure*}
Stable jetting conditions were achieved at ESRF, with an incident laser pulse energy of 2mJ and approximately 0.05 total absorbency of the laser power in water mixed with nile blue dye resulting in approximately \unit{100}{\micro J} absorbed energy per pulse. This was enough to form the repeatable jet (Fig. \ref{fig:esrf_results}a). This result is consistent with previous reported results \cite{Tagawa:2012}. The measured water jet speed is 184 m/s and the wall velocity of the laser cavity expanding wall is reaching 272 m/s. The transformation of the meniscus into the jet during first frames is also clearly visible with clear detail. Due to limited time during the EuXFEL experiment, jetting conditions were achieved. However, due to the high contrast achieved and high spatial resolution the microstructure of the laser interaction with the sample is revealed with great detail (Fig.\ref{fig:xfel_results}), which is not possible by visible light microscopy. 
To compare quantitatively the imaging performance for both experiments, we used a signal-to-noise ratio (SNR) analysis and power spectrum analysis. Detailed description about applied analysis is described in the  supplementary document. As a result, the SNR analyzed for both ESRF and EuXFEL sequences show almost two times higher values for the maximum SNR from each analyzed sequence for EuXFEL with SNR=10.69 and for ESRF with SNR=6.19 (Fig. \ref{fig:esrf_results}b,d). A more objective comparison using spectral power (Fig. \ref{fig:esrf_results}c) clearly indicates the superior performance of EuXFEL microscopy over ESRF. However stronger fluctuation in the mean values and SNR are observed for XFEL data. This is a natural behaviour of XFEL beams and is related to the origin of X-ray pulse generation using the SASE process. Such fluctuations are the reason of the lack of a procedure for correct data normalization, as every sequence and pulse has a different intensity distribution. Simple normalization of such data leads to image flickering and the increase of the standard deviation of the signal. Another contribution responsible for the increase of the standard deviation for the actual EuXFEL data is attributed to the high-frequency noise caused by the focusing KB optics. 
To remove normalization artifacts and high frequency noise for the EuXFEL data we performed an adaptive high pass filtering by subtracting a low-pass filtered (Gaussian convolution with the standard deviation $\sigma$ = 5 pixels) image from its original version. This procedure significantly suppressed spatio-temporal image flickering. Using such image processing, we visualized the velocities of breaking glass reaching 35 m/s using flow analysis based on variational optical flow methods \cite{Myagotin12}. The computed velocities provide quantitative information about complex kinematics of the burst process. The high spatial coherence of XFEL data allowed us to apply a contrast-transfer-function (CTF) based phase retrieval method utilizing the alternating direction method of multipliers (ADMM) \cite{Pablo:2017}. ADMM-CTF-based phase retrieval (fig~\ref{fig:xfel_results}). Here the data were first normalized using a set of empty beam sequences, then the ADMM-CTF phase retrieval algorithm was applied. High frequency noise introduced by this process was removed by non-local-means denoising \cite{Antoni:11}. ADMM-CTF phase retrieval of the synchrotron data failed to provide meaningful phase reconstruction, which is attributed to the weak signal. 

In conclusion, we have successfully demonstrated X-ray microscopy sampled at greater than one megahertz at EuXFEL and with the full performance of EuXFEL maximum frame-rate of 4.5 MHz will be achievable. Our model system shows a significant improvement in the contrast of the data obtained at EuXFEL compared to data obtained at ESRF. This is due to the much higher photon flux per pulse as well as the much higher spatial coherence at EuXFEL. Such performance at EuXFEL additionally allowed for a 2.5-fold increase in the spatial resolution and a significant improvement in power spectrum over entire range of frequencies, which enabled us to apply a single-distance phase retrieval algorithm. This study opens up new perspectives for imaging, especially of irreversible stochastic processes not accessible via visible light imaging or with less intense X-ray sources. At hard X-Ray MHz rate XFEL facilities this method enables the observation of stochastic object motions at high velocities on the order of m/s to several km/s. High flux per pulse at EuXFEL will enable 3D MHz rate microscopy by employing beam spliters, which is the scope of our future development. 

\section*{Funding Information}
Bundesministerium für Bildung und Forschung (BMBF) (05K18XXA), Vetenskapsrådet (VR) (2017-06719).
\section*{Acknowledgments}
We acknowledge Klaus Giewekemeyer, Luis Morillo Lopez, Cedric Michel Signe Takem, Alexis Legrand, Bradley Manning and Nadja Reimers for technical support during preparation of the EuXFEL experiment. Beamtime was granted at ESRF-beamline ID19 in the frame of proposal MI-1267.  
\newline
\newline

\section{Supplementary document}

\subsection{Experimental description}

\begin{figure}[!htp]
 \includegraphics[width = 1\textwidth]{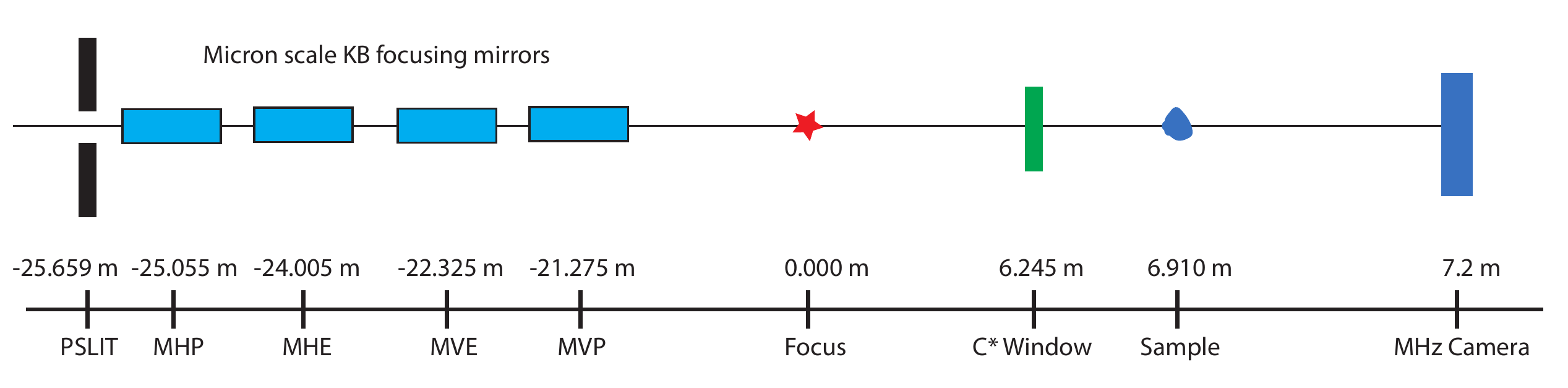}%
\caption{Optical configuration for MHz imaging at EuXFEL SPB/SFX instrument.}
\label{fig:xfel_optical_configuration}
\end{figure}

For the experiment at EuXFEL we used the SPB/SFX instrument \cite{Mancuso:2019}. For the X-ray microscopy measurements, we re-used the spent beam from the upstream interaction region that had passed through the central hole of an AGIPD detector \cite{Schwandt:2013} and was out-coupled into air via a 180 \unit{\micro\metre} thick diamond window. A simultaneous crystallography experiment and timing measurement using a photon arrival monitor \cite{Kirkwood:2019} was performed during MHz X-ray imaging. The beam from the SASE1 undulator source was delivered by the horizontal offset mirrors placed \unit{260}{\metre} downstream of the source \cite{Sinn:2019}. KB mirrors installed in the optics hutch \cite{Bean:2016} were used to focus the beam into the sample chamber with a focal spot size of \unit{3}{\micro\metre} $\times$ \unit{3}{\micro\metre}. The focus-to-detector distance was 7.2 m, the sample-to-detector distance was 0.29 m and the total air path between the diamond window and the detector was 0.96 m. The optical configuration is depicted on Fig. \ref{fig:xfel_optical_configuration}. 

A photon energy of 9.3 keV was used and the pulse train was filled with 128 X-ray pulses with a repetition rate of 1.128 MHz. The 10$\times$ magnification by the diffraction-limited microscope coupled to the 8 \textmu m thick LYSO:Ce sctintillator resulted in an effective pixel size of 3.2 \textmu m and a FOV of 1.28 mm $\times$ 0.8~mm. The Mitutoyo near-ultraviolet wavelength corrected M Plan Apo infinity corrected objective with numerical aperture 0.28 was coupled with Mitutoyo MT-L 200 mm tube lens. The spatial resolution of this imaging system was therefore limited by the effective pixel size of the FT-CMOS camera Shimadzu HPV-X2. 
The X-ray pulse energy measured with a gas detection monitor at the end of the SASE 1 tunnel was about \unit{750}{\micro\joule}. 
The sample position in the interaction region of the X-ray and visible light lasers were locked optically using two high-resolution optical microscopes. The timing alignment was performed using temporal position of the scattered radiation of the X-ray and visible laser pulses from the scattering material placed at the sample position measured using a fast silicon diode connected to oscilloscope (LeCroy WaveRunner 8404M). The time alignment of the visible light pump laser with X-ray pump laser was done by shifting the synchronized 10Hz train signal, which indicates the start of each pulse train by a delay generator (Stanford Research DG645).

\begin{figure*}[!htp]
 \includegraphics[width = 1\textwidth]{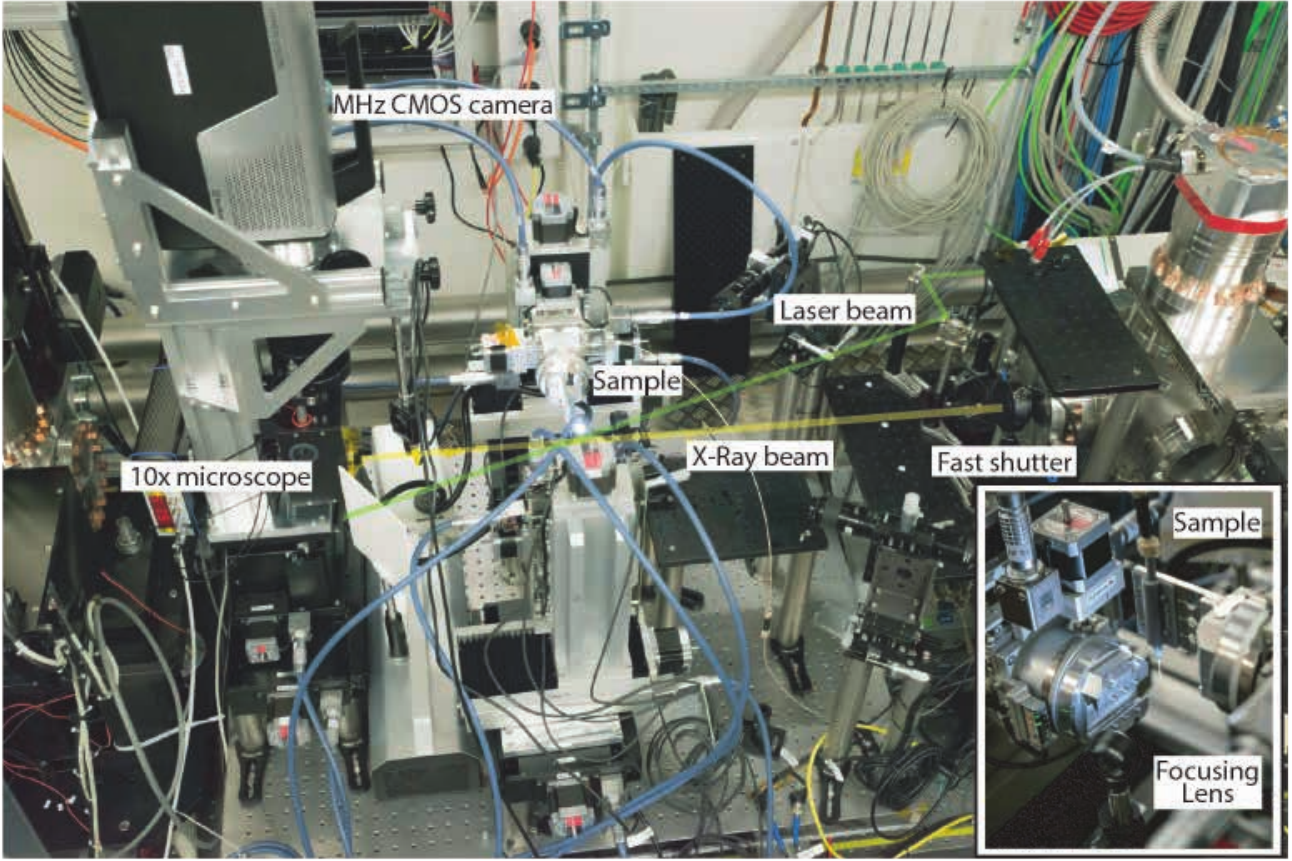}%
\caption{Experimental setup for MHz microscopy at EuXFEL SPB/SFX instrument.}
\label{fig:xfel_setup}
\end{figure*}

\begin{figure*}[!htp]
 \includegraphics[width = 1\textwidth]{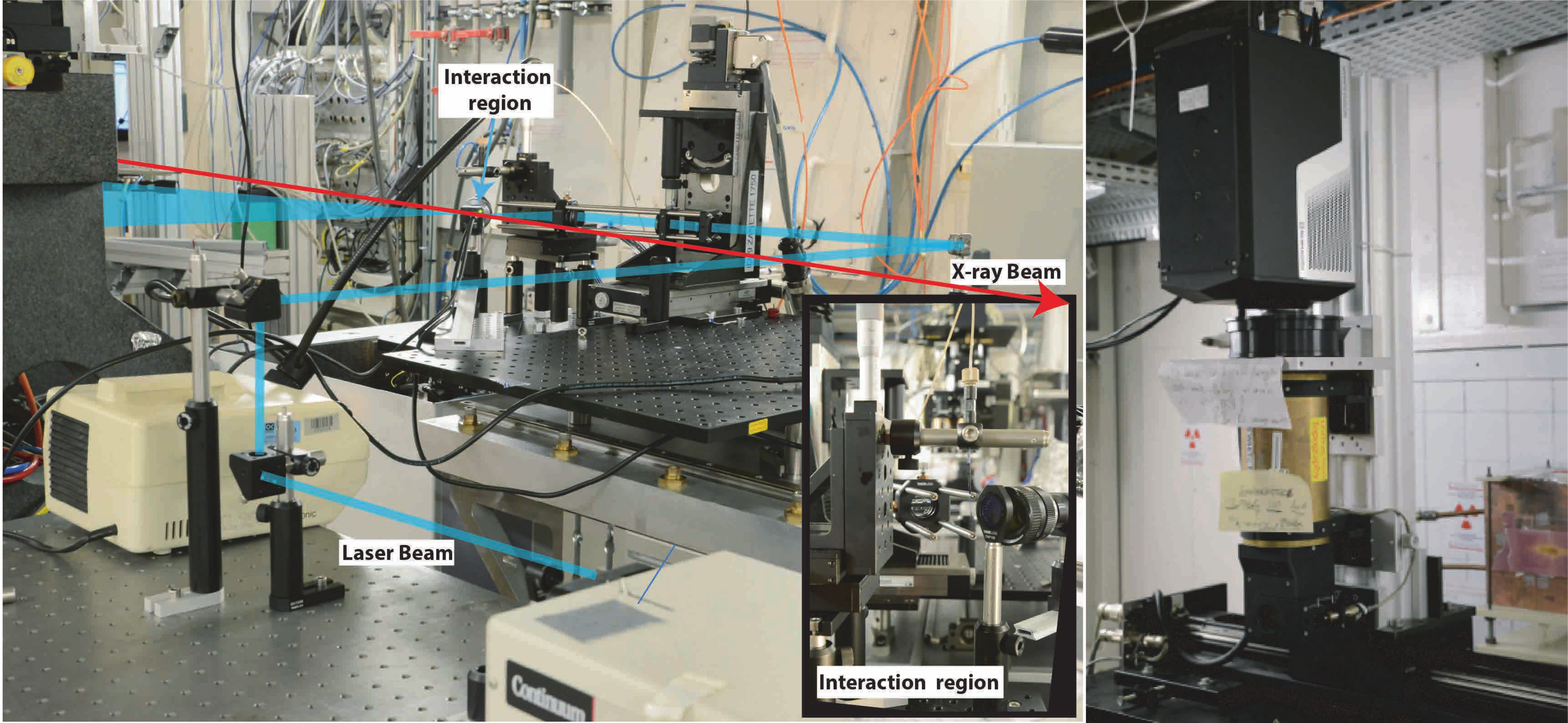}%
\caption{Experimental setup at ESRF ID19 beamline.}
\label{fig:xfel_setup}
\end{figure*}

As a sample we used a glass capillary with internal diameter of \unit{300}{\micro\metre} and wall thickness of \unit{25}{\micro\metre}. The capillary was filled with purified water and supplied via a remotely controlled syringe pump. 

For the experiment at the ESRF synchrotron we used the 16-bunch filling mode \cite{ESRF_filling_modes} providing a bunch separation of 176 ns and the MHz camera was recording every third frame with an inter-frame time of 530 ns. The harmonics of the undulator with central photon energy 32 keV were conditioned by a set of 1D compound refractive lenses to enhance the flux density at the detector. The fast camera was coupled to the microscope with 4$\times$ magnification, 0.2 numerical aperture and 250 \textmu m thick LYSO:Ce scintillator providing an effective pixel size of 8 \textmu m and a FOV of 2.0~mm $\times$ 3.0~mm. The sample-to-detector distance was 1 m. 
As a sample we used a glass capillary with internal diameter of \unit{500}{\micro\metre} and wall thickness of \unit{270}{\micro\metre}. Purified water was mixed with Nile blue dye to enhance absorption of laser power in the water. 
A pulsed Nd:YAG laser (Minilite II, Continuum) was used to generate the water jet by power absorption in the water. The laser wavelength was 532 nm with a pulse duration 3-5 ns, and the attenuated laser beam energy was 2.03 mJ with a focal spot size of \unit{300}{\micro\metre}. The 10 Hz laser flash lamp and Q-switch operation were synchronized with the RF system using the BCDU8 unit and a delay generator (DG 645, Stanford Instruments).

\subsection{Image quality comparison}

\begin{figure*}[!htp]
 \includegraphics[width = 0.90\textwidth]{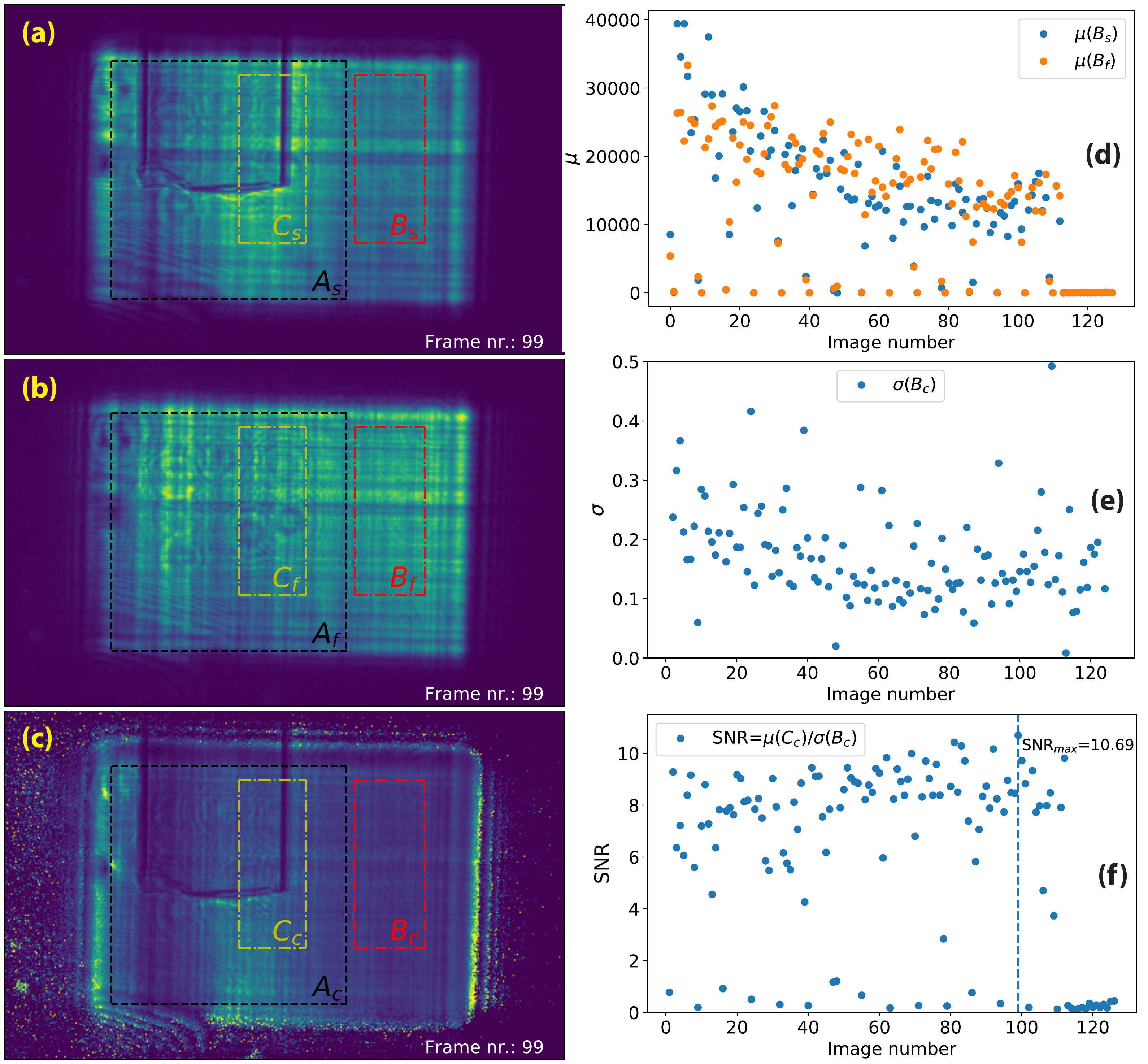}%
\caption{SNR comparison on a static sample using EuXFEL data. Signal evaluation was done by looking at the raw sequences with sample (a) and without sample (b) using the areas marked in the figures ($B_{s,f,c}$ and $C_{s,f,c}$), where the subscripts indicate: s - sequence with sample, f - flat sequence without sample and c - corrected sequence. Area $A_c$ was used to evaluate the power spectrum. Figure (d) shows the mean values for areas $B_s$ and $B_f$ over the image sequence revealing strong pulse-to-pulse intensity fluctuations. The standard deviation was evaluated on the background corrected area $B_c$ and finally the SNR showing a maximum of 10.69 was calculated by division of mean values (d) by standard deviation values (e) for a given frame in the sequence.}
\label{fig:snr_xfel}
\end{figure*}

\begin{figure*}[!htp]
 \includegraphics[width = 1\textwidth]{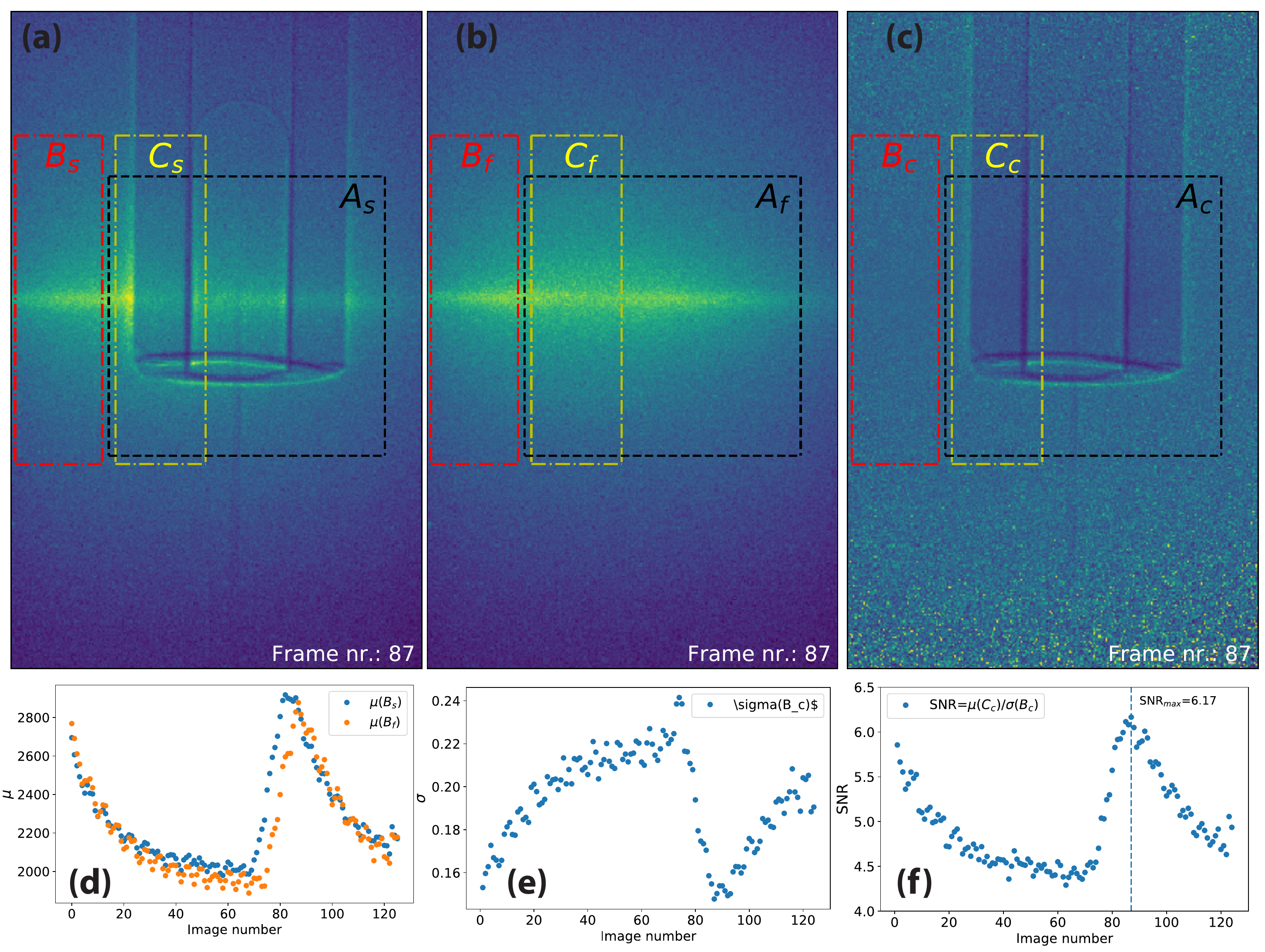}%
\caption{SNR comparison on a static sample using ESRF data. Signal evaluation was done by using the raw sequences with sample (a) and without sample (b) looking at the areas marked in the figures ($B_{s,f,c}$ and $C_{s,f,c}$). Area $A_c$ was to evaluate power spectrum. Figure (d) shows the mean values for areas $B_s$ and $B_f$ over the image sequence. The standard deviation was evaluated on the background corrected area $B_c$ and finally standard deviation showing maximum of 6.17 was calculated by division of mean values (d) by standard deviation values (e) for given frame in the sequence.}
\label{fig:snr_esrf}
\end{figure*}

To compare quantitatively the image quality between the two experiments, first the signal-to-noise ratio (SNR) was evaluated. Figures \ref{fig:snr_xfel} and \ref{fig:snr_esrf} show the results and frames used for the comparison. For each sequence frames with the sample were used and normalized by the frames without the sample. The SNR for each frame in the sequence was calculated by division of the mean values of normalized area containing the signal of the sample $C_c$ by the standard deviation of the normalized area $B_c$ without the sample. As can be clearly seen from the distribution of the mean values (Fig.\ref{fig:snr_xfel}d and Fig.\ref{fig:snr_esrf}d) the pulse-to-pulse intensity fluctuations are strong in case of EuXFEL, which is due to the origin of the photons generation at XFELs using self-amplified spontaneous emission (SASE). This makes it difficult to perform a standard normalization of image frames by the background frame and it leads to the fluctuation of the standard deviation. Such fluctuations at the synchrotron are much smaller as it is clearly seen from the mean values in Fig. \ref{fig:snr_esrf}. Another source of standard deviation increase in case of the EuXFEL data is the strong high-frequency noise caused by the beam delivery KB optics manifested as vertical and horizontal stripes. This can be in future improved by better illumination, for example by using direct beam illumination. The longer-period oscillations in mean values in Fig. \ref{fig:snr_esrf} are present due to the not exact synchronization of the camera to the master frequency of the X-ray pulses as the camera clock has a different frequency. Such oscillations are present as well in the EuXFEL data but with much smaller period resulting in only little or no illumination of certain frames in the sequence.

\begin{figure*}[!htp]
 \includegraphics[width = 1\textwidth]{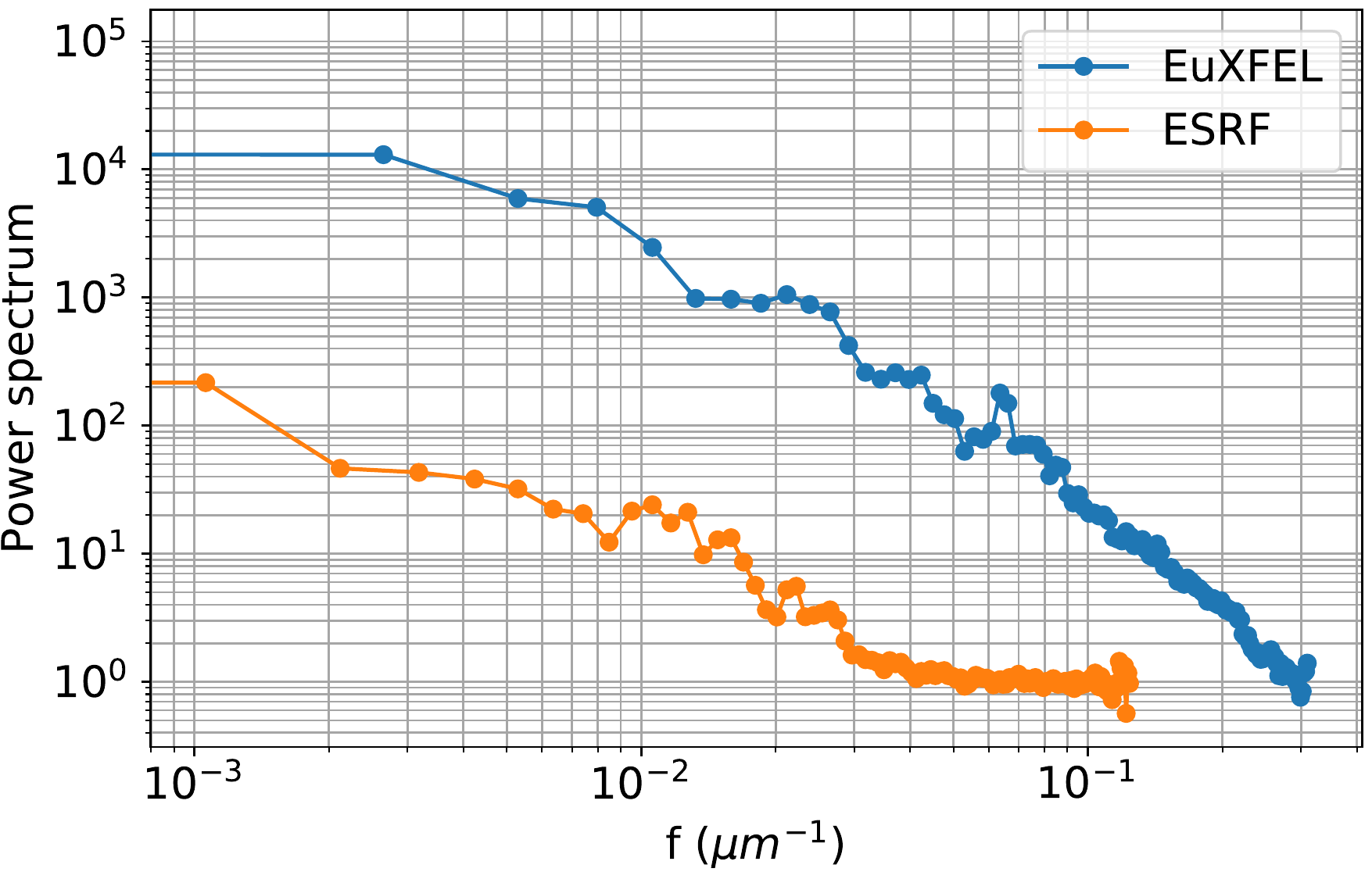}%
\caption{Power spectra of the frames with the highest SNR for the XFEL (SNR = 10.69) and the ESRF (SNR = 6.17) data. Power spectra were calculated from the normalized frames using areas $A_c$ depicted in figures \ref{fig:snr_xfel} and \ref{fig:snr_esrf}.}
\label{fig:power_spectrum}
\end{figure*}

By comparing the SNR of the two data sets, we can conclude that the EuXFEL data have a clearly higher SNR by almost a factor of two. The results from the EuXFEL also indicate that despite using a much less efficient imaging system and a higher magnification, a stronger signal is detected. This performance can unlock fast dynamics at even higher spatial resolution. However, due to the very different experimental conditions (different photon energy, magnification, scintillator thickness, optics efficiency) such comparison is not entirely correct. A more objective comparison can be achieved by looking at the power spectrum. For this purpose we selected the frames with the highest SNR for both the ESRF and the EuXFEL data and calculated the power spectrum from normalized areas $A_c$ from both data sets. Using a Fourier transform, we calculated the power of the corresponding frames as

\begin{equation}
    P(k_x,k_y) = |FFT(A_c(x,y))|^2
\end{equation}

and then integrated the spectrum in polar coordinates. The resulting graphs are shown in Fig. \ref{fig:power_spectrum}. The EuXFEL power spectrum is superior over the ESRF power spectrum as it is higher by one to almost two orders of magnitude over the entire range of spatial frequencies. 

\newpage
\bibliographystyle{unsrt}  



\end{document}